\documentclass[12pt]{article}
\usepackage{graphicx}
\usepackage{hyperref}

\def \bea{\begin{eqnarray}}
\def \beq{\begin{equation}}
\def \ca{{\cal A}}
\def \eea{\end{eqnarray}}
\def \eeq{\end{equation}}

\def \ok{\overline{K}^0}
\def \oks{\overline{K}^{*0}}

\def \stt{\sqrt{2/3}}

\def \thet{\theta_\eta}

\def \c{\circ}

\begin{document}

\rightline{EFI 11-10}
\rightline{arXiv:1104.4962}
\rightline{April 2011}

\bigskip
\centerline{\bf RELATIVE PHASES IN $D^0\to K^0 K^-\pi^+$}
\centerline{\bf AND $D^0\to\ok K^+\pi^-$ DALITZ PLOTS}
\bigskip
\centerline{Bhubanjyoti Bhattacharya and Jonathan L. Rosner}
\centerline{\it Enrico Fermi Institute and Department of Physics}
\centerline{\it University of Chicago, 5640 S. Ellis Avenue, Chicago, IL 60637}

\begin{quote}
The processes $D^0\to K^0 K^-\pi^+$ and $D^0\to\ok K^+\pi^-$ involve
intermediate vector resonances whose amplitudes and phases are related to each
other via flavor SU(3) symmetry. A closer look at Dalitz plots for these two
processes is expected to shed light on our understanding of the usefulness of
flavor SU(3) symmetry in studying charm decays. In the present work we use data
from the BaBar Collaboration's publication in 2002.  The goal is to reproduce
Dalitz plot amplitudes and phases assuming flavor SU(3) symmetry and compare
them with experiment.

While an SU(3) fit is able to account for the relative magnitudes of the
amplitudes for the decays $D^0 \to K^{*-}K^+$ and $D^0 \to K^{*+}K^-$, the
current BaBar sample (based on an integrated luminosity of only 22 fb$^{-1}$)
provides only $1$--$2 \sigma$ evidence for the decays $D^0\to K^{*0}\ok$ and
$D^0 \to \oks K^0$, with magnitudes and phases not in accord with predictions.
The CLEO collaboration could potentially produce an analysis using data with
better statistics, and an analysis based on the full BaBar sample (more than
20 times the 2002 value) should definitively settle the question.
\end{quote}

\leftline{PACS numbers:13.25.Ft, 11.30.Hv, 14.40.Lb}
\bigskip

\section{Introduction}
An important contribution to the decay processes $D^0\to 3P$, where $P$
represents a pseudoscalar meson, involves the intermediate step in which the
$D$ meson first decays into a $P$ and a vector meson ($V$). The vector meson
then decays into two pseudoscalars.  In general, in a decay with three final
$P$ states the combination of any pair of final pseudoscalars may result from
the decay of a $V$ as long as charge, isospin, strangeness, etc.\ are
conserved.  Evidence of formation of such resonances is seen in Dalitz plots as
bands of events corresponding to the invariant mass-squared of the pair of
final state $P$ mesons.  As such, they provide information about the amplitude
and phase for the process $D\to P V$. Overlapping vector resonance bands on
Dalitz plots interfere according to their relative phases.

Amplitudes and phases of $D\to P V$ decays were studied in detail using an
SU(3) flavor symmetry formalism in Ref.\ \cite{Bhattacharya:2008ke}. Relative
phase relations based on SU(3) flavor symmetry were exploited in Refs.\
\cite{Bhattacharya:2010id, Bhattacharya:2010ji, Bhattacharya:2010tg} to observe
the successes of the flavor SU(3) symmetry formalism in predicting
interferences on several $D\to 3P$ Dalitz plots. In the present article we
consider the Dalitz plots for $D^0\to K^0 K^-\pi^+$ and $D^0\to \ok K^+\pi^-$.
We extract amplitudes and phases for the relevant $D \to PV$ intermediate
processes from data published by the BaBaR collaboration \cite{Aubert:2002yc},
and compare them with theoretical predictions using flavor SU(3) symmetry.

We briefly review the flavor SU(3) symmetry technique in Sec.\ II. In Sec.\
III we quote the theoretical results for relevant $D \to PV$ processes, and
compare them with data in Sec.\ IV.  We conclude in Sec.\ V.

\section{Review of flavor SU(3) symmetry technique}

\begin{figure}
\includegraphics[width=0.48\textwidth]{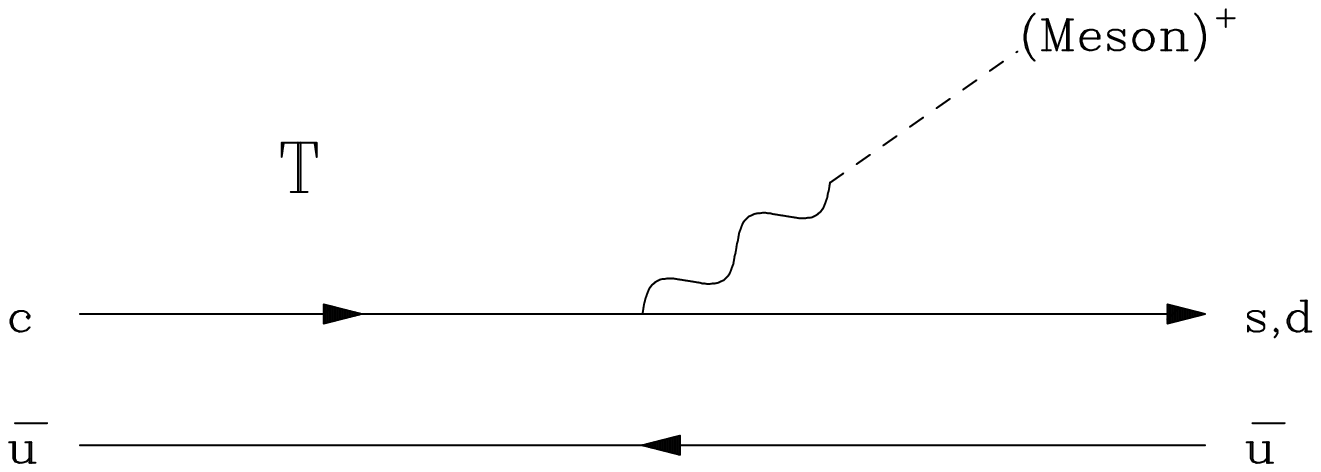}
\includegraphics[width=0.48\textwidth]{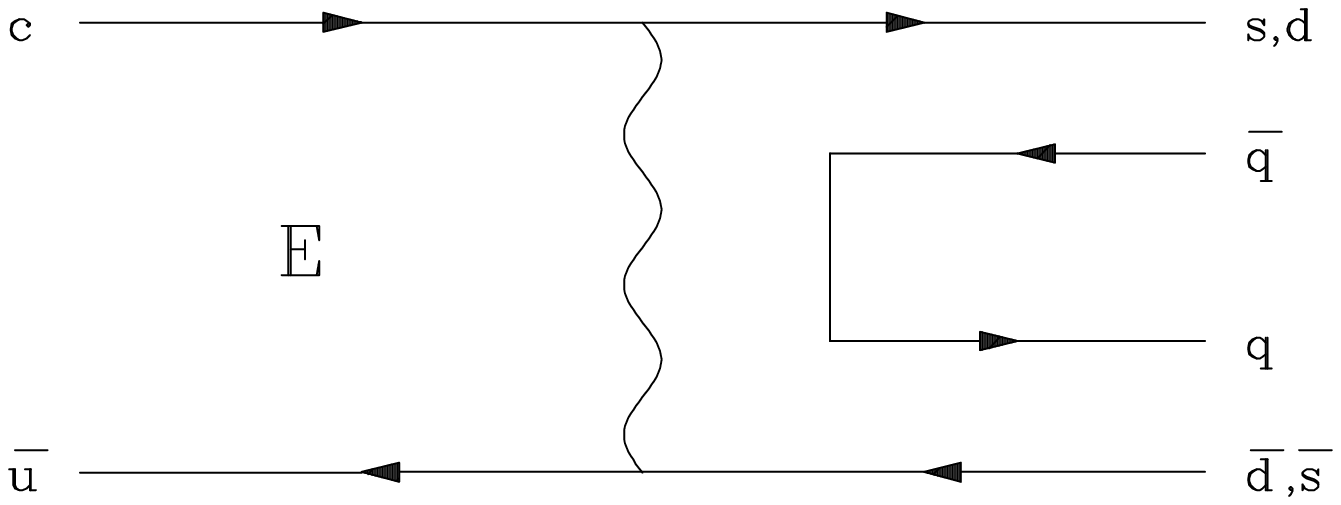}
\caption{Graphs describing tree ($T$) and exchange ($E$) amplitudes
\label{fig:TE}}
\end{figure}

The flavor symmetry approach to be used here was discussed in detail in
\cite{Bhattacharya:2008ke}. Here we recall the basic points.  We denote the
relevant Cabibbo-favored (CF) amplitudes, proportional to the product
$V_{ud}V^*_{cs}$ of Cabibbo-Kobayashi-Maskawa (CKM) factors, by amplitudes
labeled as T (``tree'') and E (``exchange''), illustrated in Fig. \ref{fig:TE}.
The singly-Cabibbo-suppressed (SCS) amplitudes, proportional to the product
$V_{us} V^*_{cs}$ or $V_{ud}V^*_{cd}$, are then obtained by using the ratio
SCS/CF $= \tan\theta_C \equiv \lambda = 0.2305$ \cite{Nakamura:2010}, with
$\theta_C$ the Cabibbo angle and signs governed by the relevant CKM factors.
The subscript $P$ or $V$ on an amplitude denotes the meson ($P$ or $V$)
containing the spectator quark in the $PV$ final state. The partial width
$\Gamma(H \to PV)$ for the decay of a heavy meson $H$ is given in terms of an
invariant amplitude $\ca$ as:
\beq
\Gamma(H \to PV) = \frac{p^{*3}}{8\pi M^2_H}|\ca|^2
\eeq
where $p^*$ is the center-of-mass (c.m.) 3-momentum of each final particle, and
$M_H$ is the mass of the decaying heavy meson. With this definition the
amplitudes $\ca$ are dimensionless.

The amplitudes $T_V$ and $E_P$ were obtained from fits to rates of CF $D \to
PV$ decays not involving $\eta$ or $\eta'$ \cite{Bhattacharya:2008ke}. To
specify the amplitudes $T_P$ and $E_V$, however, one needs information on the
$\eta - \eta'$ mixing angle ($\thet$). Table \ref{tab:tveptpev} summarizes
these results for two values $\thet = 19.5^\circ$ and $11.7^\circ$.

\begin{table}[h]
\caption{Solutions for $T_V$, $E_P$, $T_P$ and $E_V$ amplitudes in
Cabibbo-favored charmed meson decays to $PV$ final states, for $\eta$--$\eta'$
mixing angles of $\thet = 19.5^\c$ and $11.7^\circ$.
\label{tab:tveptpev}}
\begin{center}
\begin{tabular}{c c c c c} \hline \hline
 & \multicolumn{2}{c}{$\thet=19.5^\c$} & \multicolumn{2}{c}{$\thet=11.7^\c$} \\
$PV$ &  Magnitude  &  Relative  &  Magnitude  &  Relative \\
ampl.& ($10^{-6}$) & strong phase & ($10^{-6}$) & strong phase \\ \hline
$T_V$& 3.95$\pm$0.07 & Assumed 0 & \multicolumn{2}{c}{These results are}\\
$E_P$& 2.94$\pm$0.09 & $\delta_{E_PT_V} = (-93\pm3)^\circ$ & \multicolumn{2}{c}
{independent of $\thet$}\\ \hline
$T_P$ & 7.46$\pm$0.21 & Assumed 0 & 7.69$\pm$0.21 & Assumed 0 \\
$E_V$ & 2.37$\pm$0.19 &$\delta_{E_VT_V} =(-110 \pm 4)^\c$ & 1.11$\pm$0.22 &
 $\delta_{E_VT_V} =(-130 \pm 10)^\c$ \\ \hline\hline
\end{tabular}
\end{center}
\end{table}

\section{Relevant $D^0\to PV$ processes}

In Tables \ref{tab:amps19} and \ref{tab:amps11} we list the $D^0 \to PV$
amplitudes relevant in Dalitz plots of interest for $\thet = 19.5^\circ$ and
$\thet = 11.7^\circ$, respectively.  Also included are their representations
and values in terms of flavor-SU(3) amplitudes.

Flavor SU(3) symmetry requires the amplitudes $\ca(D^0\to K^{*0}\ok)$ and
$\ca(D^0 \to \oks K^0)$ to be equal in magnitude but $180^\circ$ apart in
phase. Since these two processes show up in two different Dalitz plots, this
offers us a way to check the relative amplitudes and phases within individual
Dalitz plots obtained from Dalitz plot fits. In the following section we obtain
the amplitudes and phases for the above amplitudes using Dalitz plot fit
fractions and relative phases and compare these results with theoretical
predictions using flavor SU(3) symmetry.

\begin{table}
\caption{Amplitudes for $D^0 \to PV$ decays of interest for the present
discussion (in units of $10^{-6}$). Here we have taken $\thet=19.5^\circ$.
\label{tab:amps19}}
\begin{center}
\begin{tabular}{c c c c c c c} \hline \hline
Dalitz & $D^0$ final & Amplitude & \multicolumn{4}{c}{Amplitude $A$} \\
 plot  & state  & representation & Re & Im & $|A|$ & Phase ($^\circ$) \\ \hline
$D^0 \to K^0 K^- \pi^+$&$K^{*+} K^-$ & $\lambda(T_P+E_V)$ & 1.533 & --0.513
 & 1.616 & --18.5 \\
 &$\oks K^0$ & $\lambda(E_V-E_P)$ & --0.151&  0.163 & 0.223 & 132.8 \\ \hline
$D^0 \to \ok K^+ \pi^-$&$K^{*-} K^+$ & $\lambda(T_V+E_P)$ & 0.875 & --0.677
 & 1.106 & --37.7 \\
 &$K^{*0} \ok$ & $\lambda(E_P-E_V)$ & 0.151 &--0.163 & 0.223 &--47.2 \\ \hline
\hline
\end{tabular}
\end{center}
\end{table}

\begin{table}
\caption{Amplitudes for $D^0 \to PV$ decays of interest for the present
discussion (in units of $10^{-6}$). Here we have taken $\thet=11.7^\circ$.
\label{tab:amps11}}
\begin{center}
\begin{tabular}{c c c c c c c} \hline \hline
Dalitz & $D^0$ final & Amplitude & \multicolumn{4}{c}{Amplitude $A$} \\
 plot  & state  & representation & Re & Im & $|A|$ & Phase ($^\circ$) \\ \hline
$D^0 \to K^0 K^- \pi^+$&$K^{*+} K^-$ & $\lambda(T_P+E_V)$ & 1.608 & --0.196
 & 1.620 & -- 6.9 \\
 &$\oks K^0$ & $\lambda(E_V-E_P)$ & --0.129& 0.481 & 0.498 &  105.0 \\ \hline
$D^0 \to \ok K^+ \pi^-$&$K^{*-} K^+$ & $\lambda(T_V+E_P)$ & 0.875 & --0.677
 & 1.106 & --37.7 \\
 &$K^{*0} \ok$ & $\lambda(E_P-E_V)$ & 0.129 &--0.481 & 0.498 &--75.0 \\ \hline
\hline
\end{tabular}
\end{center}
\end{table}

\section{Comparison of data with theoretical predictions}

In order to obtain amplitudes and phases for the amplitudes mentioned in the
previous section from Dalitz plot fit fractions, one needs to keep in mind that
the $D\to PV$ process is an intermediate to the complete 3 body decay $D \to
3P$. The Dalitz plot fit fractions also contain some information about the
vector meson decay and this needs to be factored out in order for any
comparison with theoretical predictions from the flavor-SU(3) method. This,
however, is fairly simple since the fraction of a vector meson's decay
amplitude to a pair of $P$ mesons can be given by the relevant isospin
Clebsch-Gordan factor.

To obtain the correct Clebsch-Gordan factor, one notes that the
spin part of the amplitude for the process $D \to RC \to ABC$ ($R$ represents
the intermediate resonance while $A$, $B$ and $C$ are the final state
pseudoscalar mesons) is proportional to the product $\vec{p}_A\cdot\vec{p}_C$
($\vec{p}_i$ is the 3-momentum of the final state particle $i$ in the rest
frame of $R$.) Since the particles $A$ and $B$ have
equal and opposite 3-momenta in the resonance rest frame, this implies then
that swapping $A$ and $B$ while calculating the amplitude would result in an
additional phase difference of $\pi$. It is thus important to know the phase
convention used to obtain the amplitudes. In the present case, due to the
absence of a stated phase convention in Ref.\ \cite{Aubert:2002yc}, we assume
a cyclic permutation convention often employed by the BaBar Collaboration
elsewhere \cite{Aubert:2008zu}.  This convention is presented in Table
\ref{tab:conv}. Using this convention one may then calculate the appropriate
isospin Clebsch-Gordan coefficients, also noted in Table \ref{tab:conv}.

The phase space factors for the two $D\to PV$ processes from each Dalitz plot
are not the same since the vector mesons involved have slightly different
masses. This very small difference, noted in Table \ref{tab:conv}, has been
neglected.

\begin{table}
\caption{Conventions for the order of two pseudoscalar mesons in vector meson
decay and associated Clebsch-Gordan factors assuming the cyclic convention
of Ref.\ \cite{Aubert:2008zu}.
\label{tab:conv}}
\begin{center}
\begin{tabular}{c c c c c c c} \hline \hline
Dalitz Plot & \multicolumn{2}{c}{Bachelor Particle} & \multicolumn{3}{c}{Vector Meson Decay} & $p^*$\\
&Meson&Index&Process&Indices&Clebsch Factor & (in MeV)\\ \hline
&$K^0$&1&$\oks\to K^-\pi^+$&23& -- $\stt$ & 605\\
$D^0\to K^0 K^-\pi^+$&$K^-$&2&$K^{*+}\to\pi^+ K^0$&31& $\stt$ & 610 \\
&$\pi^+$&3&--&--&--&--\\ \hline
&$\ok$&1&$K^{*0}\to K^+\pi^-$&23& $\stt$&605 \\
$D^0\to \ok K^+\pi^-$&$K^+$&2&$K^{*-}\to\pi^- \ok$&31& -- $\stt$&610 \\
&$\pi^-$&3&--&--&--&-- \\ \hline \hline
\end{tabular}
\end{center}
\end{table}

Using the appropriate Clebsch-Gordan coefficients we now translate the fit
parameters into amplitudes and phases that can be compared with theoretical
predictions. In Table \ref{tab:data} we quote the fit fractions and phases from
a fit to BaBaR data \cite{Aubert:2002yc} for relevant intermediate $D^0\to PV$
decays corresponding to each Dalitz plot. Fit fractions quoted in Table
\ref{tab:data} are normalized so as to represent percentage of each decay mode
in the specific Dalitz plots. This normalization is different for the two
different Dalitz plots. In order to compare amplitudes for $D\to PV$ processes
from two different Dalitz plots
it is useful to choose a universal normalization. To achieve this we
make use of the total branching fraction for the $D\to 3P$ process for each
Dalitz plot, so as to calculate the fraction of each $D\to PV$ process relative
to a common rate or amplitude.  In Table \ref{tab:data}, in addition to the
above data, we also quote the total branching fractions for the overall $D^0
\to 3 P$ process in each Dalitz plot, relative to the process $D^0\to\ok\pi^+
\pi^-$.

We make use of the BaBar data \cite{Aubert:2002yc} quoted in Table
\ref{tab:data} to calculate the relative amplitudes and phases of the relevant
$D\to PV$ decays. The magnitudes and phases of the amplitudes are obtained
relative to that of the process $D^0\to K^{*+} K^-$ with maximum amplitude.
These results are listed under the last two columns in Table \ref{tab:comp}. In
Table \ref{tab:comp} we also list the predictions of amplitudes and phases
for the same processes obtained using the flavor-SU(3)-symmetry technique.

The ratio of the amplitude $|\ca(D^0 \to K^{*-}K^+)|$ relative to
$|\ca(D^0 \to K^{*+} K^-)|$ is predicted to be equal to a corresponding ratio
of Cabibbo-favored amplitudes:
\beq
\frac{|\ca(D^0 \to K^{*-}K^+)|}{|\ca(D^0 \to K^{*+}K^-)|} =
\frac{|\ca(D^0 \to K^{*-}\pi^+)|}{|\ca(D^0 \to \rho^{*+}K^-)|}~.
\eeq
The left-hand side is $0.618^{+0.078}_{-0.089}$ as computed using the experimental
numbers from Table \ref{tab:comp}. On the other hand, the right-hand side is
$0.685 \pm 0.032$, when computed from the respective Cabibbo-favored amplitudes
\cite{Bhattacharya:2008ke}. Thus flavor SU(3) symmetry seems to be obeyed for
the dominant $VP$ sub-amplitudes in $D^0 \to K_S K^\pm \pi^\mp$.

\begin{table}
\caption{Dalitz plot fit to data from the BaBar collaboration
\cite{Aubert:2002yc}.
\label{tab:data}}
\begin{center}
\begin{tabular}{c c c c c} \hline \hline
Dalitz Plot & Branching Fraction ($\%$) & $D^0$ final
 & \multicolumn{2}{c}{Experiment} \\
 & (rel to $D^0\to\ok\pi^+\pi^-$) & state & Fit Fraction ($\%$)
 & Phase ($^\circ$) \\ \hline
$D^0\to K^0 K^-\pi^+$ & 8.32$\pm$0.29$\pm$0.56&$K^{*+} K^-$
 & 63.6$\pm$5.1$\pm$2.6&0 (fixed) \\
 & & $\oks   K^0$&0.8$\pm$0.5$\pm$0.1&175$\pm$22 \\ \hline
$D^0\to \ok K^+\pi^-$ & 5.68$\pm$0.25$\pm$0.41&$K^{*-} K^+$
 & 35.6$\pm$7.7$\pm$2.3&0 (fixed) \\
 & & $K^{*0} \ok$&2.8$\pm$1.4$\pm$0.5&--126$\pm$19 \\ \hline \hline
\end{tabular}
\end{center}
\end{table}

\begin{table}
\caption{Amplitudes for $D^0\to PV$ decays from Dalitz plots of interest for
the present discussion (in units of $10^{-6}$). Here we have taken $\thet =
19.5^\circ$, and $\lambda = 0.2305$ \cite{Nakamura:2010}. The experimental
amplitudes have been taken from BaBar data \cite{Aubert:2002yc} and have a
normalization such that the largest amplitude is fixed to 1.  The phases in
lines 2 and 3 of the last column have been flipped by $180^\circ$ in
comparison with those listed in lines 2 and 3 of Table \ref{tab:data} to
correspond to the negative signs of the Clebsch-Gordan coefficients in
Table \ref{tab:conv}.
\label{tab:comp}}
\begin{center}
\begin{tabular}{c c c c c c} \hline \hline
$D^0$ final & Amplitude & \multicolumn{2}{c}{Theory}
 & \multicolumn{2}{c}{Experiment} \\
state & Representation & Amplitude & Phase ($^\circ$) & Amplitude
 & Phase ($^\circ$) \\ \hline
$K^{*+} K^-$ &$\lambda(T_P+E_V)$&1.616$\pm$0.060&--18.5$\pm$1.6&1 (fixed)
 & 0 (fixed)\\
$\oks K^0$ &$\lambda(E_V-E_P)$&0.223$\pm$0.050&132.8$\pm$13.0
 & 0.112$^{+0.032}_{-0.045}$&--5$\pm$22\\
$K^{*-} K^+$ &$\lambda(T_V+E_P)$&1.106$\pm$0.033&--37.7$\pm$1.5
 & 0.618$^{+0.078}_{-0.089}$& 180 (fixed)\\
$K^{*0} \ok$ &$\lambda(E_P-E_V)$&0.223$\pm$0.050&--47.2$\pm$13.0
 & 0.173$^{+0.042}_{-0.057}$&--126$\pm$19 \\
\hline \hline
\end{tabular}
\end{center}
\end{table}

Flavor SU(3) predicts equal magnitudes for the much smaller amplitudes $\ca(D^0
\to \oks K^0)$ and $\ca(D^0 \to K^{*0} \ok)$. The central values of the
magnitudes obtained from the experimental fit are respectively smaller and
larger than the theoretical prediction:
\beq
\frac{|\ca(D^0 \to \oks K^0)|}{|\ca(D^0 \to K^{*+}K^-)|} = 0.112^{+0.032}
_{-0.045} ~~({\rm expt.})~~vs.\ 0.138 \pm 0.033 ~~({\rm theory})~;\label{eqn:3}
\eeq
\beq
\frac{|\ca(D^0 \to \ok K^{*0})|}{|\ca(D^0 \to K^{*+}K^-)|} = 0.173^{+0.042}
_{-0.057} ~~({\rm expt.})~~vs.\ 0.138 \pm 0.033 ~~({\rm theory})~.\label{eqn:4}
\eeq

In the above equations we only quote the $1\sigma$ error bars for the ratios.
Although the probability distribution functions for the input branching
fractions may be taken to be Gaussian, the probability distribution for the
amplitude ratios ({\rm expt.}) are quite different. In fact one may check
that the ratios ({\rm expt.}) in Eqns.\ (\ref{eqn:3}) and (\ref{eqn:4}) are
consistent with zero at the $1.55 \sigma$ and $1.81 \sigma$ levels,
respectively.

Given the large experimental errors, the discrepancies in magnitudes from the
flavor-SU(3) predictions are not yet evidence for violation of this
symmetry.  However, relative phases of the relevant amplitudes obtained from
theory and experiment are not in agreement, the discrepancies being
$(156 \pm 22)^\circ$ for $D^0 \to K^0 K^- \pi^+$ and $(64 \pm 19)^\circ$ for
$D^0 \to \ok K^+ \pi^-$, quoting only the experimental error.  Similar
conclusions follow from the predictions for $\thet = 11.7^\circ$.  This could
arise from a misunderstood convention for the vector meson decay, as explained
in the previous section, or could signal a breakdown of the flavor-SU(3)
approach.  If the latter, it would be the first such instance, as earlier
analyses \cite{Bhattacharya:2010id, Bhattacharya:2010ji, Bhattacharya:2010tg}
reproduced such relative phases successfully. One may also argue that the 
experimental relative phases and the error bars on them are meaningless, 
since the corresponding amplitude ratios are consistent with zero. In that 
case one needs a larger data sample to identify the correct relative phases 
and corresponding error bars.

We conclude for a number of reasons that it is premature to declare flavor
SU(3) invalid for the decays in question:  (1) The BaBar $K^{*0} \ok$ and
$\oks K^0$ amplitudes are marginal and have not yet been confirmed by any other
experiment; (2) BaBar's phase convention has not been explicitly stated and,
though requested, has not been made available; (3) results from CLEO with a
larger data sample will soon be available; and (4) BaBar's total sample is more
than 20 times as large and an updated analysis would provide much more
convincing statistics.

\section{Conclusions}

Flavor SU(3) has had notable success in predicting relative phases of
Dalitz plot amplitudes in charm \cite{Bhattacharya:2010id, Bhattacharya:2010ji,
Bhattacharya:2010tg} and beauty \cite{Gronau:2010kq} decays.  Nevertheless, the
pattern of interference between vector meson resonance bands in $D^0 \to K^0
K^- \pi^+$ and $D^0 \to \ok K^+ \pi^-$ Dalitz plots, based on analysis of a
small fraction of currently available data \cite{Aubert:2002yc}, is at odds
with SU(3) predictions.  Notably, (1) the relative phase between $K^{*+} K^-$
and $\oks K^0$ bands in $D^0 \to K^0 K^- \pi^+$ differs from the SU(3)
prediction by $(156 \pm 22)^\circ$, where we quote only the experimental error;
(2) the relative phase between vector meson resonance bands in $D^0 \to \ok K^+
\pi^-$ differs from the SU(3) prediction by $(64 \pm 19)^\circ$; (3) the ratio
$|\ca(D^0 \to \oks K^0)/\ca(D^0 \to K^{*0} \ok)|$, predicted to be 1, is
$0.64 \pm 0.27$.

In view of the facts that the BaBar analysis \cite{Aubert:2002yc} employed an
integrated luminosity of only 22 fb$^{-1}$, whereas a data sample of more
than twenty times that is now available, and that the CLEO Collaboration also
has a large sample of such events, this Dalitz plot would appear to be a prime
target for re-analysis.

\section*{Acknowledgements}

We thank Brian Meadows and Guy Wilkinson for helpful discussions. This work was
supported in part by the United States Department of Energy through
Grant No.\ DE FG02 90ER40560.

\end{document}